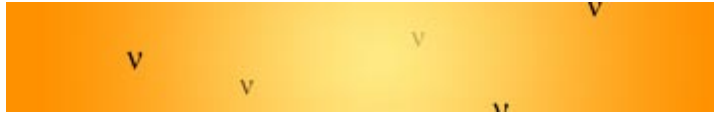

# Solving the Mystery of the Missing Neutrinos

by John N. Bahcall

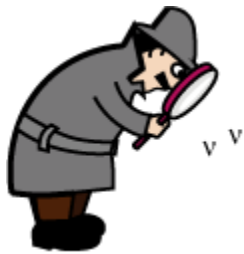

The three years 2001 to 2003 were the golden years of solar neutrino research. In this period, scientists solved a mystery with which they had been struggling for four decades. The solution turned out to be important for both physics and for astronomy. In this article, I tell the story of those fabulous three years.[1]

The first two sections summarize the solar neutrino mystery and present the solution that was found in the past three years. The next two sections describe what the solution means for physics and for astronomy. The following sections outline what is left to do in solar neutrino research and give my personal view of why it took more than thirty years to solve the mystery of the missing neutrinos. The last section provides a retrospective impression of the solution.

## The Mystery

### The Crime Scene

During the first half of the twentieth century, scientists became convinced that the Sun shines by converting, deep in its interior, hydrogen into helium. According to this theory, four hydrogen nuclei called protons (p) are changed in the solar interior into a helium nucleus ($^4$He), two anti-electrons ($e^+$, positively charged electrons), and two elusive and mysterious particles called neutrinos ($\nu_e$). This process of nuclear conversion, or nuclear fusion, is believed to be responsible for sunshine and therefore for all life on Earth. The conversion process, which involves many different nuclear reactions, can be written schematically as:

$$4p \rightarrow {}^4He + 2e^+ + 2\nu_e \qquad (1).$$

Two neutrinos are produced each time the fusion reaction (1) occurs. Since four protons are heavier than a helium nucleus, two positive electrons and two neutrinos, reaction (1) releases a lot of energy to the Sun that ultimately reaches the Earth as sunlight. The reaction occurs very frequently. Neutrinos escape easily from the Sun and their energy does not appear as solar heat or sunlight. Sometimes neutrinos are produced with relatively low energies and the Sun gets a lot of heat. Sometimes neutrinos are produced with higher energies and the Sun gets less energy.

The neutrinos in equation (1) and the illustration below are the focus of the mystery that we explore in this article.

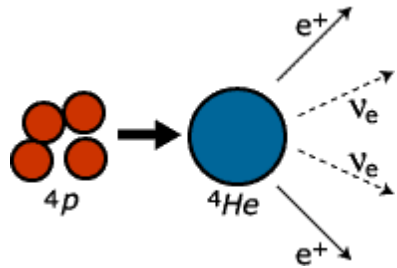

Neutrinos have zero electric charge, interact very rarely with matter, and—according to the textbook version of the standard model of particle physics—are massless. About 100 billion neutrinos from the Sun pass through your thumbnail every second, but you do not feel them because they interact so rarely and so weakly with matter. Neutrinos are practically indestructible; almost nothing happens to them. For every hundred billion solar neutrinos that pass through the Earth, only about one interacts at all with the stuff of which the Earth is made. Because they interact so rarely, neutrinos can escape easily from the solar interior where they are created and bring direct information about the solar fusion reactions to us on Earth. There are three known types of neutrinos. Nuclear fusion in the Sun produces only neutrinos that are associated with electrons, the so-called electron neutrinos ($v_e$). The two other types of neutrinos, muon neutrinos ($v_\mu$) and tau neutrinos ($v_\tau$), are produced, for example, in laboratory accelerators or in exploding stars, together with heavier versions of the electron, the particles muon ($\mu$) and tau ($\tau$).

**Neutrinos Are Missing**

In 1964, following the pioneering work of Raymond Davis Jr., he and and I proposed an experiment to test whether converting hydrogen nuclei to helium nuclei in the Sun is indeed the source of sunlight, as indicated by equation (1).

I calculated with my colleagues the number of neutrinos of different energies that the Sun produces using a detailed computer model of the Sun and also calculated the number of radioactive argon atoms ($^{37}$Ar) these solar neutrinos would produce in a large tank of chlorine-based cleaning fluid ($C_2Cl_4$). Although the idea seemed quixotic to many experts, Ray was sure that he could extract the predicted number of a few atoms of $^{37}$Ar per month out of a tank of cleaning fluid that is about the size of a large swimming pool.

The first results of Ray's experiment were announced in 1968. He detected only about one third as many radioactive argon atoms as were predicted. This discrepancy between the number of predicted neutrinos and the number Ray measured soon became known as "The Solar Neutrino Problem" or, in more popular contexts, "The Mystery of the Missing Neutrinos."

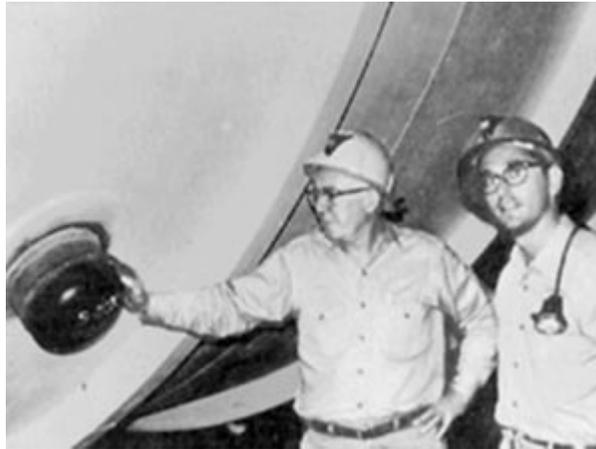

Raymond David Jr. (left) and John Bahcall in miner's clothing and protective hats. The photograph was taken in 1967 about a mile underground in the Homestake Gold Mine in Lead, South Dakota, USA. Davis is pictured showing Bahcall his newly constructed steel ank (6 meters in diameter, 15 meters long), which contained a large amount of cleaning fluid (40,000 liters) and was used to capture neutrinos from the Sun.
Photo: Courtesy of Raymond Davis, Jr. and John Bahcall

### Possible Explanations

Three classes of explanation were suggested to solve the mystery. First, perhaps the theoretical calculations were wrong. This could happen in two ways. Either the predicted number of neutrinos was incorrect or the calculated production rate of argon atoms was not right. Second, perhaps Ray's experiment was wrong. Third, and this was the most daring and least discussed possibility, maybe physicists did not understand how neutrinos behave when they travel astronomical distances.

The theoretical calculations were refined and checked many times over the next two decades by me and by different researchers. The data used in the calculations were improved and the predictions became more precise. No significant error was found in the computer model of the Sun or in my calculation of the probability of Ray's tank capturing neutrinos. Similarly, Ray increased the sensitivity of his experiment. He also carried out a number of different tests of his technique in order to make sure that he was not overlooking some neutrinos. No significant error was found in the measurement. The discrepancy between theory and experiment persisted.

What about the third possible explanation, new physics? Already in 1969, Bruno Pontecorvo and Vladimir Gribov of the Soviet Union proposed the third explanation listed above, namely, that neutrinos behave differently than physicists had assumed. Very few physicists took the idea seriously at the time it was first proposed, but the evidence favoring this possibility increased with time.

### Evidence Favors New Physics

In 1989, twenty-one years after the first experimental results were published, a Japanese-American experimental collaboration reported the results of an attempt to "solve" the solar neutrino problem. The new experimental group called Kamiokande (led by Masatoshi Koshiba and Yoji Totsuka) used a large detector of pure water to measure the rate at which electrons in the water scattered the highest-energy neutrinos emitted from the Sun. The water detector was very sensitive, but only to high-energy neutrinos that are produced by a rare nuclear reaction (involving the decay of the nucleus $^8$B) in the solar energy production cycle. The original Davis experiment with chlorine was primarily, but not exclusively, sensitive to the same high-energy neutrinos.

The Kamiokande experiment confirmed that the number of neutrino events that were observed was less than predicted by the theoretical model of the Sun and by the textbook description of neutrinos. But, the discrepancy in the water detector was somewhat less severe than observed in the chlorine detector of Ray Davis.

In the following decade, three new solar neutrino experiments deepened the mystery of the missing neutrinos. Experiments in Italy and Russia used massive detectors containing gallium to show that lower energy neutrinos were also apparently missing. These experiments were called GALLEX (led by Till Kirsten of Heidelberg, Germany) and SAGE (led by Vladimir Gavrin of Moscow, Russia). The fact that GALLEX and SAGE were sensitive to lower energy neutrinos was very important since I believed I could calculate more accurately the number of low energy neutrinos than the number of higher energy neutrinos. In addition, a much larger version of the Japanese water detector, called Super-Kamiokande (led by Totsuka and Yochiro Suzuki), made more precise measurements of the higher energy neutrinos and confirmed the original deficit of higher energy neutrinos found by the chlorine and Kamiokande experiments. So both high and low energy neutrinos were missing, although not in the same proportions.

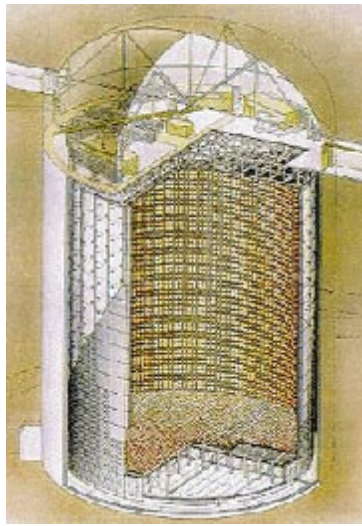

The Super-Kamiokande Detector, University of Tokyo. The detector consists of an inner volume and an outer volume which contain 32,000 and 18,000 tons of pure water, respectively. The outer volume shields the inner volume in which neutrino interactions are studied. The inner volume is surrounded by 11,000 photomultiplier tubes that detect pale blue Cherenkov light emitted when electrons are struck by neutrinos.
Drawing: Courtesy of Kamioka Observatory, ICRR, University of Tokyo

Evidence obtained during this decade indicated that something must happen to the neutrinos on their way to detectors on Earth from the interior of the Sun. In 1990, Hans Bethe and I pointed out that new neutrino physics, beyond what was contained in the standard particle physics textbooks, was required to reconcile the results of the Davis chlorine experiment and the Japanese-American water experiment. Our conclusion was based upon an analysis of the relative sensitivity of the chlorine and the water experiments to neutrino number and neutrino energy. The newer solar neutrino experiments in Italy and in Russia increased the difficulty of explaining the neutrino data without invoking new physics.

New evidence also showed that the solar model predictions were reliable. In 1997, precise measurements were made of the sound speed throughout the solar interior using periodic fluctuations observed in ordinary light from the surface of the Sun. The measured sound speeds agreed to a precision of 0.1% with the sound speeds calculated for our theoretical model of the Sun. These measurements suggested to astronomers that the theoretical model of the Sun was so accurate that the model must also predict correctly the number of solar neutrinos.

The last decade of the twentieth century provided strong evidence that a better theory of fundamental physics was required to solve the mystery of the missing neutrinos. But, we still needed to find the smoking gun.

### The Solution

On June 18, 2001 at 12:15 PM (eastern daylight time) a collaboration of Canadian, American, and British scientists made a dramatic announcement: they had solved the solar neutrino mystery. The international collaboration (led by Arthur McDonald of Ontario, Canada) reported the first solar neutrino results obtained with a detector of 1,000 tons of heavy water[2] ($D_2O$). The new detector, located in a nickel mine in Sudbury, Ontario in Canada, was able to study in a different way the same higher-energy solar neutrinos that had been investigated previously in Japan with the Kamiokande and Super-Kamiokande ordinary-water detectors. The Canadian detector is called SNO for Solar Neutrino Observatory.

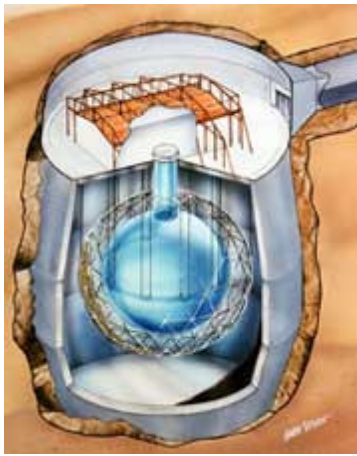

Artist's drawing showing cutaway of the Sudbury Solar Neutrino Observatory, encased in its housing and submerged in a mine. The inner detector contains 1,000 tons of heavy water and is surrounded by a stainless steel structure carrying about 10,000 photomultiplier tubes. The outer, barrel-shaped cavity (22 meters in diameter and 34 meters in height) is filled with purified ordinary water to provide support and to shield against particles other than neutrinos.
Drawing: Copyright© Garth Tietien, 1991

### The Definitive Experiments

For their first measurements, the SNO collaboration used the heavy-water detector in a mode that is sensitive only to electron neutrinos. The SNO scientists observed approximately one-third as many electron neutrinos as the standard computer model of the Sun predicted were created in the solar interior. The Super-Kamiokande detector, which is primarily sensitive to electron neutrinos but has some sensitivity to other neutrino types, observed about half as many events as were expected.

If the standard model of particle physics was right, the fraction measured by SNO and the fraction measured by Super-Kamiokande should be the same. All the neutrinos should be electron neutrinos. The fractions were different. The standard textbook model of particle physics was wrong.

Combining the SNO and the Super-Kamiokande measurements, the SNO collaboration determined the total number of solar neutrinos of all types (electron, muon, and tau) as well as the number of just electron neutrinos. The total number of neutrinos of all types agrees with the number predicted by the computer model of the Sun. Electron neutrinos constitute about a third of the total number of neutrinos.

The smoking gun was discovered. The smoking gun is the difference between the total number of neutrinos and the number of only electron neutrinos. The missing neutrinos were actually present, but in the form of the more difficult to detect muon and tau neutrinos.

The epochal results announced in June 2001 were confirmed by subsequent experiments. The SNO collaboration made unique new measurements in which the total number of high energy neutrinos of all types was observed in the heavy water detector. These results from the SNO measurements alone show that most of the neutrinos produced in the interior of the Sun, all of which are electron neutrinos when they are produced, are changed into muon and tau neutrinos by the time they reach the Earth.

The measurement of the total number of neutrinos in the SNO detector provided the fingerprints on the smoking gun.

These revolutionary results were verified independently in an extraordinary tour-de-force by a Japanese-American experimental collaboration, Kamland, which studied, instead of solar neutrinos, anti-neutrinos emitted by nuclear power reactors in Japan and in neighboring countries. The collaboration (led by Atsuto Suzuki of Sendai, Japan) observed a deficit in the detected number of anti-neutrinos from the nuclear power reactors. A deficit had been predicted for the Kamland experiment based upon the solar model calculations, the solar neutrino measurements, and a theoretical model of neutrino behavior that explained why the previous calculations and measurements seemed to be in disagreement. The Kamland measurements significantly improved our knowledge of the parameters that characterize neutrinos.

### Where Did the Missing Neutrinos Go?

The solution of the mystery of the missing solar neutrinos is that neutrinos are not, in fact, missing. The previously uncounted neutrinos are changed from electron neutrinos into muon and tau neutrinos that are more difficult to detect. The muon and tau neutrinos were not detected by the Davis experiment with chlorine; they were not detected by the gallium experiments in Russia and in Italy; and they were not detected by the first SNO measurement. This lack of sensitivity to muon and tau neutrinos is the reason that these experiments seemed to suggest that most of the expected solar neutrinos were missing. On the other hand, the Kamiokande and Super-Kamiokande water experiments in Japan and the later SNO heavy water experiments had some sensitivity to muon and tau neutrinos in addition to their primary sensitivity to electron neutrinos. These water experiments revealed therefore larger fractions of the predicted solar neutrinos.

**What Does All This Mean for Physics?**

**What Is Wrong with Neutrinos?**

Solar neutrinos have a multiple personality disorder. They are created as electron neutrinos in the Sun, but on the way to the Earth they change their type. For neutrinos, the origin of the personality disorder is a quantum mechanical process, called "neutrino oscillations."

Pontecorvo and Gribov had the right idea in 1969. Lower energy solar neutrinos switch from electron neutrino to another type as they travel in the vacuum from the Sun to the Earth. The process can go back and forth between different types. The number of personality changes, or oscillations, depends upon the neutrino energy. At higher neutrino energies, the process of oscillation is enhanced by interactions with electrons in the Sun or in the Earth. Stas Mikheyev, Alexei Smirnov, and Lincoln Wolfenstein first proposed that interactions with electrons in the Sun could exacerbate the personality disorder of neutrinos, i.e., the presence of matter could cause the neutrinos to oscillate more vigorously between different types.

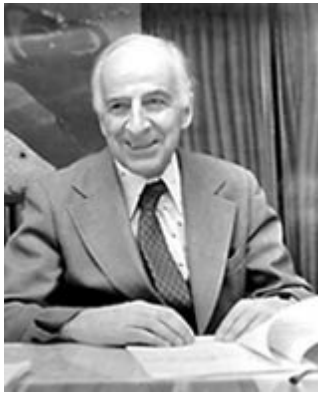

Bruno Pontecorvo in his office at the Joint Institute for Nuclear Physics, Dubna, Russia in 1983. Pontecorvo was discussing physics with his collaborator Samoil Bilenky. Later that afternoon, Pontecorvo celebrated his 70th birthday with a party.
Photo: Courtesy of Samoil Bilenky and John Bahcall

Even before the SNO measurement in 2001, phenomenological analyses of all the solar neutrino experimental data suggested with rather high confidence that some new physics was occurring. The preferred neutrino parameters from these pre-SNO analyses agreed with the parameters that were selected later with higher confidence by the SNO and Super-Kamiokande results. But, the smoking gun was missing.

The SNO and Super-Kamiokande results taken together were equivalent to finding a smoking gun, because they referred to the same high-energy solar neutrinos and because the experiments used techniques that were familiar to many physicists. Also, both experiments included many checks on their measurements.

**What Is Wrong with the Standard Model of Particle Physics?**

The standard model of particle physics assumes that neutrinos are massless. In order for neutrino oscillations to occur, some neutrinos must have masses. Therefore, the standard model of particle physics must be revised.

The simplest model that fits all the neutrino data implies that the mass of the electron neutrino is about 100 million times smaller than the mass of the electron. But, the available data are not yet sufficiently definitive to rule out all but one possible solution. When we finally have a unique solution, the values of the different neutrino masses may be clues that lead to understanding physics beyond the standard model of particle physics.

There are two equivalent descriptions of neutrinos, one that is expressed in terms of the masses of the neutrinos and one that is expressed in terms of the particles with which the neutrinos are associated (electron neutrinos with electrons, muon neutrinos with muon particles, or tau neutrinos with tau particles). The relations between the mass description and the associated-particle description involve certain constants, called "mixing angles," whose values are potentially important clues that may help lead to an improved theory of how elementary particles behave.

Solar neutrino research shows that neutrinos can change their personalities or types. The mathematical description of this malady determines quantities that we hope will be useful clues in the search for a more general theory of how fundamental particles behave.

### What Does All This Mean for Astronomy?

The total number of neutrinos observed in the SNO and Super-Kamiokande experiments agrees with the number calculated using the standard computer model of the Sun. This shows that we understand how the Sun shines, the original question that initiated the field of solar neutrino research. The solution of the mystery of the missing neutrinos is an important triumph for astronomy. The standard solar model predictions are vindicated; the standard model of particle physics must be revised. Four decades ago, when the first solar neutrino experiment was proposed, no one would have guessed that this turn of events would be the outcome.

In order to predict correctly the number of neutrinos produced by nuclear reactions in the Sun, many complicated phenomena must be understood in detail. For example, one must understand a smorgasbord of nuclear reactions at energies where measurements are difficult. One must understand the transport of energy at very high temperatures and densities. One must understand the state of the solar matter in conditions that cannot be studied directly on Earth. The temperature at the center of the Sun is about 50,000 times higher than the temperature on Earth on a sunny day and the density in the center of the Sun is about a hundred times the density of water. One must measure the abundances of the heavy elements on the surface of the Sun and then understand how these abundances change as one goes deeper into the Sun. All of these and many more details must be understood and calculated accurately.

The predicted number of high-energy solar neutrinos can be shown by a quantum mechanical calculation to depend sensitively on the central temperature of the Sun. A 1% error in the temperature corresponds to about a 30% error in the predicted number of neutrinos; a 3% error in the temperature results in a factor of two error in the neutrinos. The physical reason for this great sensitivity is that the energy of the charged particles that must collide to produce the high-energy neutrinos is small compared to their mutual electrical repulsion. Only a small fraction of the nuclear collisions in the Sun succeed in overcoming this repulsion and causing fusion; this fraction is very sensitive to the temperature. Despite this great sensitivity to temperature, the theoretical model of the Sun is sufficiently accurate to predict correctly the number of neutrinos.

The research efforts of thousands of researchers in institutions distributed throughout the world have been necessary to achieve the required precision. As a result of this community effort over the past four decades, we now have confidence in our understanding of how stars shine. We can use this knowledge to interpret observations of distant galaxies that also contain stars. We can use the theory of how stars shine and evolve to learn more about the evolution of the universe.

### What Is Left To Do?

The chlorine and gallium detectors do not measure the energy of neutrino events. Only the water detectors (Kamiokande, Super-Kamiokande, and SNO) provide specific information about the energies of the solar neutrinos that are observed. However, the water detectors are sensitive only to higher energy neutrinos (with energies > 5 million electron volts).

The standard computer model of the Sun predicts that most solar neutrinos have energies that are below the detection thresholds for the water detectors. If the standard solar model is correct, water detectors are sensitive to only about 0.01% of the neutrinos the Sun emits. The remaining 99.99% must be observed in the future with new detectors that are sensitive to relatively low energies.

The Sun is the only star close enough to the Earth for us to observe the neutrinos produced by nuclear fusion reactions. It is important to observe the abundant low-energy solar neutrinos in order to test more precisely the theory of stellar evolution. We believe we can calculate the expected number of low energy neutrinos more accurately than we can calculate the number of high-energy neutrinos. Therefore, an accurate measurement of the number of low energy neutrinos will be a critical test of the degree of accuracy of our solar theory. There may still be surprises.

At lower energies (< 2 million electron volts), we believe that the theory of Pontecorvo and Gribov describes well the conversion in vacuum of electron neutrinos into neutrinos of other types. At higher energies, we think that interactions with electrons, as suggested by Mikheyev, Smirnov, and Wolfenstein, are required in order to understand the enhanced conversion of electron neutrinos into other types of neutrinos. We need new experiments at low energies to test for, and understand quantitatively, the change in the conversion mechanism from the process operating at high energies to the process that is most important at low energies.

Solar neutrino experiments at low energies can also provide refined measurements of the parameters that describe neutrino oscillations.

We can use neutrinos to measure the total radiant luminosity of the Sun. The present estimate of the total luminosity uses only the particles of light, called photons. If the only source of radiant energy is nuclear fusion reactions as described by the equation shown in the beginning of the article, then the two measurements (with light and with neutrinos) will agree. We expect agreement based upon our current understanding of how the sun shines. But, if there is another source of energy—some process that we do not yet know about—then the measurements with neutrinos and with light may differ significantly. That would be a revolutionary discovery.

**Why Did It Take So Long?**

The mystery of the missing solar neutrinos was first recognized in 1968. The number of neutrino events observed by Ray Davis in his detector was much less than the predicted value. But, it was not until 2001 that most physicists were convinced that the origin of the solar neutrino mystery was an inadequacy in the standard model of particle physics rather than a failure of the standard theoretical model of how the Sun shines.

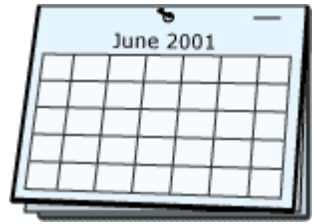

Why did it take so long for most physicists to be convinced that the particle theory was wrong and not the astrophysics?

Let's first hear in their own words what some of the most prominent physicists have said about the missing neutrinos. In 1967, two years before his epochal paper with Gribov on solar neutrino oscillations was published, Bruno Pontecorvo wrote:

"Unfortunately, the weight of the various thermonuclear reactions in the sun, and the central temperature of the sun are insufficiently well known in order to allow a useful comparison of expected and observed solar neutrinos..."

In other words, the uncertainties in the solar model are so large that they prevent a useful interpretation of solar neutrino measurements. Bruno Pontecorvo's view was echoed more than two decades later when in 1990 Howard Georgi and Michael Luke wrote as the opening sentences in a paper on possible particle physics effects in solar neutrino experiments:

"Most likely, the solar neutrino problem has nothing to do with particle physics. It is a great triumph that astrophysicists are able to predict the number of $^8$B neutrinos to within a factor of 2 or 3..."

C. N. Yang stated on October 11, 2002, a few days after the awarding of the Nobel Prize in Physics to Ray Davis and Masatoshi Koshiba for the first cosmic detection of neutrinos, that:

"I did not believe in neutrino oscillations even after Davis' painstaking work and Bahcall's careful analysis. The oscillations were, I believed, uncalled for."

Sidney Drell wrote in a personal letter of explanation to me in January 2003 that "… the success of the Standard Model [of particle physics] was too dear to give up."

The standard model of particle physics is a beautiful theory that has been tested and found to make correct predictions for thousands of laboratory experiments. The standard solar model, on the other hand, involves complicated physics in unfamiliar conditions and had not previously been tested to high precision. Moreover, the predictions of the standard solar model depend sensitively on details of the model, such as the central temperature. No wonder it took scientists a long time to blame the standard model of particle physics rather than the standard model of the Sun.

### An Astonishing Community Achievement

I am astonished when I look back on what has been accomplished in the field of solar neutrino research over the past four decades. Working together, an international community of thousands of physicists, chemists, astronomers, and engineers has shown that counting radioactive atoms in a swimming pool full of cleaning fluid in a deep mine on Earth can tell us important things about the center of the Sun and about the properties of exotic fundamental particles called neutrinos. If I had not lived through the solar neutrino saga, I would not have believed it was possible.

[1]This article is self-contained and can be read independently, but it is a sequel to the article "How the Sun Shines" by J.N. Bahcall that was presented on the Nobel e-Museum Web site, in June 2000. In the three years following the publication of the original article, a flood of confirmatory experimental data has been obtained. These new data provide by themselves a fascinating story that is presented here.

[2]Heavy water is chemically similar to ordinary water. However, the hydrogen in heavy water has a nucleus consisting of a proton and a neutron and is called deuterium. For ordinary water, the hydrogen has a nucleus that contains only a proton (and no neutron).

Printout of http://www.nobel.se/physics/articles/bahcall/